\documentstyle[aps]{revtex}

\begin{document}
\title{Two-Body Density Matrix for Closed $s-d$ Shell Nuclei}
\author{S.S.Dimitrova, D.N.Kadrev, A.N.Antonov,\ M.V.Stoitsov}
\address{Institute of Nuclear Research and Nuclear Energy, Bulgarian\\
Academy of Sciences, Sofia 1784, Bulgaria}
\maketitle

\begin{abstract}
The two-body density matrix for $^{4}He,^{16}O$ and $^{40}Ca$ within the
Low-order approximation of the Jastrow correlation method is considered.
Closed analytical expressions for the two-body density matrix, the center of
mass and relative local densities and momentum distributions are presented.
The effects of the short-range correlations on the two-body nuclear
characteristics are investigated.
\end{abstract}

\section{ Introduction}

Nowadays two-body knock-out reactions such as $(\gamma ,NN)$ \cite{1,2} and $%
(e,e^{\prime }NN)$ \cite{3,4,5,6} are intensively studied in order to
extract some information about the nucleon-nucleon short-range correlations
(SRC) in nuclei. In particular, the cross section for the two-nucleon
emission processes is generally related to the two-body spectral function at
least in PWIA \cite{book}. The first calculations of the two-nucleon
spectral function of $^{16}O$ have already been performed \cite{9} by
treating the long-range correlations within a Dressed RPA approach and
including the SRC in terms of the defect functions emerging as a solution of
the Bethe--Goldstone equation for finite nuclei. The results have been
successfully applied for calculating the\thinspace $(e,e^{\prime }pp)$
cross-sections \cite{6}.

Due to the complexity of the problem, however, it is highly desirable to
reach a theoretical description of the two-body knock-out reactions directly
in terms of the nuclear ground state (e.g., the ground-state two-body
density matrix) without the necessity to deal with the two-body spectral
function which is an enormously more complicated object due to the presence
of various excited states of the system. Making use of series of more or
less controlled approximations, people usually try to incorporate in this
context simplified expressions or combinations of physical quantities such
as the two-particle spectroscopic factors and overlap functions \cite{pacati}%
, relative and center-of-mass pair momentum distributions, combined two-body
momentum distributions \cite{orlandini} and generalized momentum
distributions (see e.g. \cite{clark}). Realistic but still elaborated
description of the ground state properties of nuclei up to $^{16}O$ is known
from the Variational Monte-Carlo calculations \cite{spw,car,pwp} and
recently from the Green's Function Monte-Carlo method performed for nuclei
with mass number $A\leq 7$ \cite{GFMC}. In any case, one still needs a
simple and possibly analytical model description of the ground state
two-body density matrix in order to clarify the two-body nuclear
characteristics as well as the extent to which they are affected by the
nucleon-nucleon correlations.

The situation is quite similar to the theoretical description of
one-particle removal processes in terms of the nuclear ground state
characteristics. At the beginning, it has been demonstrated \cite{r8} that
the knowledge of the ground-state one-body density matrix of the target
nucleus is sufficient to restore the single-particle overlap functions,
spectroscopic factors and separation energies associated with the bound $%
(A-1)$-particle eigenstates. Then, quantitative estimates have been obtained 
\cite{sad} within a simple and analytical one-body density matrix model \cite
{r9} which takes into account the short--range correlations in terms of the
Jastrow correlation method. Plausible conclusions have been made for the
properties of singe-particle overlap functions in comparison with the
associated shell model orbitals and the natural orbitals \cite{r9,r12,r15}
which are of frequent interest in this context \cite{r7,r13}. The resulting
overlap functions and spectroscopic factors have been used to analyze
differential cross sections of $(p,d)$ reactions and single-particle
momentum distributions in $(e,e^{\prime }p)$ reactions \cite{r28}. Finally,
more sophisticated representations of the one-body density matrix \cite
{20,21,22} have been used for extracting overlap functions and spectroscopic
factors further applied for analyzing $(p,d)$ reaction cross-sections \cite
{23}. Thus, the resulting comparative study \cite{23} has clarified the
impact of the different types of nucleon-nucleon correlations on the
one-particle removal reaction cross-sections.

Recently, similar restoration procedure has been proposed in \cite{2ovf}
connecting the two-nucleon overlap functions associated with the bound
states of the $(A-2)$ particle system with the asymptotic behavior of the
ground state two-body density matrix of the $A$-particle system. This makes
it possible to calculate, at least in principle, the two-body overlap
functions, spectroscopic factors and separation energies on the basis of
realistic representations for the ground state two-body density matrix.
Again, as a first step, one clearly needs a simple and analytical
representation of the two-body density matrix which adequately reflects the
properties of the realistic nuclear ground state and also takes into account
the short--range correlations in nuclei.

The present paper suggests such a simple and analytical representation of
the ground state two-body density matrix derived within the Low-order
approximation of the Jastrow correlation method. The associated two-particle
nuclear characteristics for the closed $s-d$ shell nuclei $^{4}He,^{16}O$
and $^{40}Ca$ are analyzed. A comparison with the realistic Variational
Monte-Carlo calculations is also made. It justifies the physical meaning of
the approximation as a tool for analyzing the impact of the nuclear SRC in a
simple way. The analytical expressions obtained can facilitate the
explanation of most of the questions arising from the interpretation of
physical properties which are important in treating the two-particle
emission processes in nuclei.

The paper is organized as follows. The general definitions of the two-body
density matrix and the associated nuclear characteristics in coordinate and
momentum representation are introduced in Sec.II. The analytical expressions
derived within the Low-order approximation of the Jastrow correlation method
are given in Sec.III. Closed analytical expressions for the two-body nucleon
momentum and density distributions are collected in Sec.IV. Results and
discussions are given in Sec.V, while the conclusions are summarized in
Sec.VI. The Appendix contains the coefficients entering the analytical
expressions for $^{4}He$ and $^{16}O$ nuclei.

\section{Two-body density matrix}

\subsection{Definitions and properties}

The physical antisymmetric state of a system of $A$ identical fermions $\Psi
^{A}$ normalized to unity defines a set of density matrices of order $%
p=1,2,...,A$

\begin{equation}
\rho ^{(p)}(x_{1},x_{2},...,x_{p};x_{1}^{\prime },x_{2}^{\prime
},...,x_{p}^{\prime })=<\Psi |a^{\dagger }(x_{1})a^{\dagger
}(x_{2})...a^{\dagger }(x_{p})a(x_{1}^{\prime })a(x_{2}^{\prime
})...a(x_{p}^{\prime })|\Psi >\;,  \label{pbdm}
\end{equation}
where $a^{\dagger }(x_{i})$ and $a(x_{i})$ stand for creation and
annihilation operators for a nucleon at position $x_{i}$, which includes the
spatial coordinate ${\bf r}_{i}$, the spin $s_{i}$ and the isospin $\tau
_{i} $. In particular, the one- and two-body density matrices are defined in
coordinate space as:

\begin{equation}
\rho ^{(1)}(x,x^{\prime })=\langle {\Psi }^{(A)}|a^{+}(x)\,a(x^{\prime })|{%
\Psi }^{(A)}\rangle \,\,,  \label{obdm1}
\end{equation}

and

\begin{equation}
\rho ^{(2)}(x_{1},x_{2};x_{1}^{\prime },x_{2}^{\prime })=\langle {\Psi }%
^{(A)}|a^{+}(x_{1})\,a^{+}(x_{2})\,a(x_{1}^{\prime })\,a(x_{2}^{\prime })|{%
\Psi }^{(A)}\rangle \ ,  \label{tbdm1}
\end{equation}
$\,\,$respectively. From these defining equations one can easily recognize
many of the properties of the density matrices. They are Hermitian 
\begin{equation}
\rho ^{(1)}(x,x^{\prime })\equiv \rho ^{(1)*}(x^{\prime };x)\;,\;\;\rho
^{(2)}(x_{1},x_{2};x_{1}^{\prime },x_{2}^{\prime })=\rho
^{(2)*}(x_{1}^{\prime },x_{2}^{\prime };x_{1},x_{2})\;,  \label{herm}
\end{equation}
and trace--normalized to the number of particles and of pairs of particles:

\begin{equation}
{\it Tr}\,\rho ^{(1)}=\int \rho (x)\,dx\;=\;A\,\,,  \label{norm1}
\end{equation}
\begin{equation}
{\it Tr}\,\rho ^{(2)}=\frac{1}{2}\,\int \rho
^{(2)}(x_{1},x_{2})\,dx_{1}\,dx_{2}\;=\;\frac{A(A-1)}{2}\,\,,  \label{norm2}
\end{equation}
with diagonal symmetric elements 
\begin{equation}
\rho (x)=\rho ^{(1)}(x,x)\;,\;\;\rho ^{(2)}(x_{1},x_{2})=\;\rho
^{(2)}(x_{1}x_{2};x_{1}x_{2})\;.  \label{diag}
\end{equation}
In addition, the two--body density matrix $\rho ^{(2)}$ is antisymmetric in
each set of indices, e.g.,

\begin{equation}
\rho ^{(2)}(x_{1}x_{2};x_{1}^{\prime }x_{2}^{\prime })=-\rho
^{(2)}(x_{2}x_{1};x_{1}^{\prime }x_{2}^{\prime })\;,  \label{anti}
\end{equation}
so that its diagonal elements vanish identically if both coordinates are
equal, i.e., $\rho ^{(2)}(x_{1},x_{1})=0$.

The one- and two-body density matrices are related by the formula 
\begin{equation}
\int \rho ^{(2)}(x_{1}x_{2};x_{1}^{\prime }x_{2})\;dx_{2}=\frac{A-1}{2}%
\;\rho ^{(1)}(x_{1},x_{1}^{\prime })\;,  \label{rec2}
\end{equation}
and both can be presented in the momentum space using the Fourier transforms:

\begin{equation}
n^{(1)}(k;k^{\prime })=\int \rho (x,x^{\prime })\ {\it \exp }\left[ {\it i}%
\left( {\bf k}.{\bf r}-{\bf k}^{\prime }.{\bf r}^{\prime }\right) \right] d%
{\bf r}d{\bf r}^{\prime }\;  \label{obdmk}
\end{equation}

\begin{equation}
n^{(2)}(k_{1},k_{2};k_{1}^{\prime },k_{2}^{\prime })=\int \rho
^{(2)}(x_{1},x_{2};x_{1}^{\prime },x_{2}^{\prime })\ {\it \exp }\left[ {\it i%
}\left( {\bf k}_{1}.{\bf r}_{1}+{\bf k}_{2}.{\bf r}_{2}-{\bf k}_{1}^{\prime
}.{\bf r}_{1}^{\prime }-{\bf k}_{2}^{\prime }.{\bf r}_{2}^{\prime }\right)
\right] \;d{\bf r}_{1}d{\bf r}_{2}d{\bf r}_{1}^{\prime }d{\bf r}_{2}^{\prime
}\;,  \label{tbdmk}
\end{equation}
where $k_{i}$ stands for the momentum ${\bf k}_{i}$, spin $s_{i}$ and
isospin $\tau _{i}$ of the $i-$th particle. Relations similar to eqs.(\ref
{herm})-(\ref{anti}) held for the one- and two-body density matrices in the
momentum space as well.

\subsection{Two-body nuclear characteristics}

Typical ground state quantities of interest one usually considers are the
local density

\begin{equation}
\rho ({\bf r)\equiv \ }\rho ({\bf r,r)=\ }\sum_{s\tau }\rho (x{\bf ,}x{\bf ),%
}  \label{locden}
\end{equation}
the associated elastic form-factor

\begin{equation}
F({\bf q)\ =\ }\frac{1}{A}\int \rho ({\bf r)\ }{\it \exp }[i{\bf q.r]\ }d%
{\bf r}  \label{ff}
\end{equation}
and the nucleon momentum distribution

\begin{equation}
n({\bf k)\equiv \ }n({\bf k,k)=}\sum_{s\tau }n(k{\bf ,}k{\bf )}
\label{nucmom}
\end{equation}
obtained after spin (and/or isospin) summation of the diagonal elements of
the one-body density matrix in coordinate and momentum representation.

The two-particle emission experiments however require some knowledge of
physical quantities associated with the two-body density matrix. For
example, the diagonal elements of the two-body density matrix $\rho ^{(2)}$
in coordinate space, eq.(\ref{tbdm1}), define the center-of-mass pair local
density distribution: 
\begin{equation}
{\bf \;\;}\rho ^{(2)}({\bf R)=}\int \rho ^{(2)}({\bf R+s/}2{\bf ,R-s/}2{\bf )%
}d{\bf r\ }  \label{denrelR}
\end{equation}
and the relative local density distribution: 
\begin{equation}
\rho ^{(2)}({\bf s)=}\int \rho ^{(2)}({\bf R+s/}2{\bf ,R-s/}2{\bf )}d{\bf R\ 
}  \label{denrelr}
\end{equation}
while the diagonal elements in momentum space eq.(\ref{tbdmk}) define the
associated center-of-mass and relative pair momentum distributions:

\begin{equation}
n^{(2)}({\bf K)=}\int n^{(2)}({\bf K/}2{\bf +k,K/}2{\bf -k)}d{\bf k\ }
\label{denrelK}
\end{equation}
and

\begin{equation}
n^{(2)}({\bf k)=}\int n^{(2)}({\bf K/}2{\bf +k,K/}2{\bf -k)}d{\bf K\;,}
\label{denrelk}
\end{equation}
respectively.

The physical meaning of $\rho ^{(2)}({\bf s)}$ and $n^{(2)}({\bf k)}$ is the
probability to find two particles displaced of a certain relative distance $%
{\bf s=r}_{1}{\bf -r}_{2}$ or moving with relative momentum ${\bf k=(k}_{1}%
{\bf -k}_{2})/2$, respectively, while $\rho ^{(2)}({\bf R)}$ and $n^{(2)}(%
{\bf K)}$ represents the probability to find a pair of particles with
center-of-mass coordinate ${\bf R=}\left( {\bf r}_{1}+{\bf r}_{2}\right) 
{\bf /}2$ or center-of-mass momentum ${\bf K=k}_{1}{\bf +k}_{2}$,
respectively.

\section{Analytical expressions for the two-body density matrix}

\subsection{Mean-field approximation}

The mean--field approximation to the nuclear ground state of an $A$ particle
system is represented by a single Slater determinant 
\begin{equation}
\Phi _{SD}^{A}(x_{1},x_{2},\ldots ,x_{A})=\frac{1}{\sqrt{A!}}\det \left|
\varphi _{i}(x_{j})\right| \;,  \label{slater}
\end{equation}
where the orthonormalized set of single--particle functions $\varphi
_{i}(x)=\varphi _{i}({\bf r,}s,\tau )$ is emerging from some kind of shell
model or self--consistent mean--field calculations. The ground state $\Phi
_{SD}^{A}$ incorporates two kinds of correlations:\ (1) Pauli correlations
associated with the antisymmetric properties of $\Phi _{SD}^{A}$ and (2) the
correlations among the nucleons forming the nuclear mean field that
determines the particular form of the single particle states $\varphi
_{i}(x) $.

The following expressions for the one- and two-body density matrices are
well known from the mean--field theory

\begin{equation}
\rho _{SD}(x,x^{\prime })=\sum\limits_{i=1}^{A}\varphi _{i}^{\ast
}(x)\varphi _{i}(x^{\prime })\;,  \label{ro1sd}
\end{equation}

\begin{equation}
\rho _{SD}^{(2)}(x_{1}x_{2};x_{1}^{\prime }x_{2}^{\prime
})=\sum\limits_{i,j=1}^{A}\varphi _{ij}^{*}(x_{1}x_{2})\varphi
_{ij}(x_{1}^{\prime }x_{2}^{\prime })\,\ ,  \label{ro2sd}
\end{equation}
where the antisymmetric uncorrelated two-body wave functions are used 
\begin{equation}
\varphi _{ij}(x_{1}x_{2})=\frac{1}{\sqrt{2}}\,\left[ \varphi
_{i}^{*}(x_{1})\varphi _{j}(x_{2})-\varphi _{j}^{*}(x_{1})\varphi
_{i}(x_{2})\right] \;.  \label{gemsd}
\end{equation}

In order to obtain analytical expressions which will allow us to compute $%
\rho _{SD}$ and $\rho _{SD}^{(2)}$ in a direct way we further assume all
states belonging to the uncorrelated Fermi sea as represented by harmonic
oscillator single-particle wave functions $\varphi _{i}^{HO}(x)$ which
depend on the harmonic-oscillator length $\alpha $, heaving the same values
for protons and neutrons. In particular, these are the states $1s$ for $%
^{4}He$, $1s$ and $1p$ for $^{16}O$ and $1s,1p,1d$ and $2s$ for $^{40}Ca$.
Because we are interested in spin (and/or isospin) free quantities like
total center of mass and relative coordinate and momentum distributions we
consider only the matrix elements which are fully diagonal in spin and
isospin variables. Under these assumptions eq.(\ref{ro2sd}) has closed
analytical form: 
\begin{equation}
\rho _{SD}^{(2)}({\bf r}_{1},s_{1},\tau _{1};{\bf r}_{2},s_{2},\tau _{2};%
{\bf r}_{3},s_{1},\tau _{1};{\bf r}_{4},s_{2},\tau _{2})=\frac{1}{2}\ [\rho (%
{\bf r}_{1},{\bf r}_{3})\rho ({\bf r}_{2},{\bf r}_{4})-\delta _{\tau
_{1}\tau _{2}}\delta _{s_{1}s_{2}}\rho ({\bf r}_{1},{\bf r}_{2})\rho ({\bf r}%
_{3},{\bf r}_{4})],
\end{equation}
since the spin and isospin free one-body density matrix 
\begin{equation}
{\bf \ }\rho ({\bf r}_{1},{\bf r}_{2}{\bf )=\ }\sum_{s\tau }\rho _{SD}({\bf r%
}_{1},s,\tau ;{\bf r}_{2},s,\tau )
\end{equation}
is an explicit product of exponent and polynomial factors depending only on
the scaled coordinates ${\bf x}_{i}=\alpha {\bf r}_{i}$:

\begin{equation}
\rho ({\bf r}_{1},{\bf r}_{2})={\frac{\alpha ^{3}}{\pi ^{3/2}}}\,{\it exp}[{-%
\frac{{\bf x}_{1}^{2}+{\bf x}_{2}^{2}}{2}}]\;P_{SD}({\bf x}_{1}{,{\bf x}}%
_{2})\ \ ,  \label{ro2sdp1}
\end{equation}
where 
\begin{equation}
\begin{array}{lll}
P_{SD}({\bf x}_{1}{,{\bf x}}_{2}) & =1 & \text{for\ }\ ^{4}He\ , \\ 
&  &  \\ 
& ={\left( 1+2\,{x_{12}}\right) } & \text{for}\ \ ^{16}O\ , \\ 
&  &  \\ 
& =\frac{1}{2}\ \left[ 5+4x_{12}-2\ (x_{1}^{2}-2x_{12}+x_{2}^{2})\right] & 
\text{for\ \ }^{40}Ca
\end{array}
\label{ro2sdp2}
\end{equation}
and ${x_{ij}\equiv \ }{\bf x}_{i}{\bf .x}_{j}=\alpha ^{2}r_{i}r_{j}\cos
\theta _{ij}$, $\theta _{ij}$ being the angle between radius vectors ${\bf r}%
_{i}$ and ${\bf r}_{j}$.

\subsection{Jastrow correlations method}

In the present paper we consider the nucleon-nucleon SRC within the Jastrow
correlation method \cite{jas,gau,boh,dal}. This method incorporates the
nucleon-nucleon SRC in terms of the wave function ansatz:

\begin{equation}
\Psi ^{(A)}({\bf r}_{1},{\bf r}_{2},\ldots ,{\bf r}_{A})=(C_{A})^{-1/2}%
\prod_{1\leq i<j\leq A}f(\mid {\bf r}_{i}-{\bf r}_{j}\mid )\ \Phi _{SD}^{A}(%
{\bf r}_{1},{\bf r}_{2},\ldots ,{\bf r}_{A}),  \label{jas}
\end{equation}
where $\Phi _{SD}^{A}$ is a single Slater determinant, $f({\bf r)}$ is a
correlation factor which goes to unity for large values of ${\bf r}$ and $%
C_{A}$ is a normalization constant. Except for systems containing very small
number of particles it is impossible to calculate the one- and two-body
density matrices using the Jastrow ansatz eq.(\ref{jas}). One usually
applies a perturbation expansion in terms of linked diagrams \cite{gau}. The
so-called Low-order approximation (LOA) keeps all terms up to second order
in $h=f-1$ and first order in $g=f^{2}-1$ in such a way that the
normalization of the density matrices is ensured order by order \cite{gau}.
In particular, the resulting two-body density matrix is of the form \cite
{boh,dal}:

\begin{equation}
\begin{tabular}{ll}
$\rho _{LOA}^{(2)}(x_{1},x_{2};x_{3},x_{4})$ & $=\rho
_{SD}^{(2)}(1234)+\left[ \;f_{13}^{*}\;f_{24}-1\right] \;\rho
_{SD}^{(2)}(1234)+{\int dx_{5}\left[ \left( f_{15}^{*}f_{35}-1\right)
+\left( f_{25}^{*}f_{45}-1\right) \right] }$ \\ 
\  & \  \\ 
\  & ${\times }\left[ {\,\rho _{SD}(x_{1};x_{3})\rho _{SD}^{(2)}(2545)+\rho
_{SD}(x_{1};x_{4})\rho _{SD}^{(2)}(2553)+\rho _{SD}(x_{1};x_{5})\rho
_{SD}^{(2)}(2534)}\right] $ \\ 
\  & \  \\ 
\  & $+{\int \int dx_{5}dx_{6}\left[ f_{56}^{*}f_{56}-1\right] }\left\{ {%
\rho _{SD}^{(2)}(2665)\rho _{SD}^{(2)}(1534)}+{\rho _{SD}(x_{1};x_{5})\rho
_{SD}(x_{2};x_{3})\rho _{SD}^{(2)}(5646)}\right. $ \\ 
\  & \  \\ 
\  & $+\left. {\rho _{SD}(x_{1};x_{5})\rho _{SD}(x_{2};x_{4})\rho
_{SD}^{(2)}(5663)+\rho _{SD}(x_{1};x_{5})\rho _{SD}(x_{2};x_{6})\rho
_{SD}^{(2)}(5634)}\right\} ,$%
\end{tabular}
\label{ro2low}
\end{equation}
where $f_{ij}\equiv f(|{\bf r}_{i}-{\bf r}_{j}|)$, $\rho
_{SD}^{(2)}(1234)\equiv \rho _{SD}^{(2)}(x_{1},x_{2};x_{3},x_{4})$ and the
integration over $x_{i}$ means summation over the spin and isospin variables
and integration over the spacial coordinate.

It is clearly seen from eq.(\ref{ro2low}) that $\rho _{LOA}^{(2)}$ generally
depends only on two ingredients, the correlation function $f(|{\bf r}_{i}-%
{\bf r}_{j}|)$ and the two-body density matrix in its mean-filed
approximation $\rho _{SD}^{(2)}$ , eq. (\ref{ro2sd}). In the previous eqs.(%
\ref{ro2sdp1},\ref{ro2sdp2}) we have already derived $\rho _{SD}^{(2)}$ in
closed analytical form in terms of harmonic-oscillator single-particle wave
functions. In order to find closed analytical expression for $\rho
_{LOA}^{(2)}$ , eq.(\ref{ro2low}), we further assume that the correlation
factor $f({\bf r)}$ is state-independent and has simple gaussian form:

\begin{equation}
f({\bf r)\equiv \ }f(r)\ {\bf =1-c\ }{\it \exp }(-\beta ^{2}r^{2})\ ,
\label{fcorr}
\end{equation}
where the correlation parameter $\beta $ controls the healing distance,
while the parameter $c$ accounts for the strength of the SRC. Under these
additional assumptions, performing explicitly the integrations entering eq.(%
\ref{ro2low}), the diagonal part in the spin and isospin variables of $\rho
_{LOA}^{(2)}$ transforms to pure algebraic expression: 
\begin{equation}
\begin{tabular}{ll}
$\rho _{LOA}^{(2)}({\bf r}_{1},s_{1},\tau _{1};{\bf r}_{2},s_{2},\tau _{2};%
{\bf r}_{3},s_{1},\tau _{1};{\bf r}_{4},s_{2},\tau _{2})=$ & $\rho
_{SD}^{(2)}+\rho _{A}^{(2)}+\rho _{B}^{(2)}+\rho _{C}^{(2)}\;,$%
\end{tabular}
\label{rabc}
\end{equation}
where the first term $\rho _{SD}^{(2)}$ is already defined by eq.(\ref
{ro2sdp1},\ref{ro2sdp2}). The sum of this term $\rho _{SD}^{(2)}$ with the
next one

\begin{equation}
\begin{tabular}{ll}
$\rho _{A}^{(2)}=$ & $\frac{1}{2}\left\{ c^{2}{\it exp}[{-({\bf z}_{1}-{\bf z%
}_{2})^{2}-({\bf z}_{3}-{\bf z}_{4})^{2}}]-c{\it exp}[{-({\bf z}_{1}-{\bf z}%
_{2})^{2}}]-c{\it exp}[{-({\bf z}_{3}-{\bf z}_{4})^{2}}]\right\} $ \\ 
\  & \  \\ 
\  & $\times \;\left\{ \rho ({\bf r}_{1},{\bf r}_{3})\rho ({\bf r}_{2},{\bf r%
}_{4})-\delta _{\tau _{1}\tau _{2}}\delta _{s_{1}s_{2}}\rho ({\bf r}_{1},%
{\bf r}_{2})\rho ({\bf r}_{3},{\bf r}_{4})\right\} \;,$%
\end{tabular}
\end{equation}
where ${\bf z}_{i}=\beta {\bf r}_{i}$ is often referred to as a first order
approximation to the Jastrow two-body density matrix \cite{ripka}. The main
disadvantage of this approximation is that it does not satisfy the
normalization condition eq.(\ref{norm2}) and thus has restricted physical
significance. The third term $\rho _{B}^{(2)}$ in eq.(\ref{rabc}) has the
form: 
\begin{equation}
\begin{tabular}{ll}
$\rho _{B}^{(2)}$ & $=\frac{\ c^{2}}{2}\left\{ 4\left[ \rho ({\bf r}_{1},%
{\bf r}_{3})\,\rho ({\bf r}_{2},{\bf r}_{4})-\delta _{\tau _{1}\tau
_{2}}\delta _{s_{1}s_{2}}\rho ({\bf r}_{1},{\bf r}_{4})\,\rho ({\bf r}_{2},%
{\bf r}_{3})\right] \left( I_{13}+I_{24}\right) -\rho ({\bf r}_{1},{\bf r}%
_{3})\,\left( I_{1324}+I_{2424}\right) \right. $ \\ 
\  & \  \\ 
\  & $-\left. \rho ({\bf r}_{2},{\bf r}_{4})\,\left( I_{1313}+I_{2}\right)
+\delta _{\tau _{1}\tau _{2}}\delta _{s_{1}s_{2}}\,\left[ \rho ({\bf r}_{2},%
{\bf r}_{3})\,\left( I_{1314}+I_{2414}\right) +\;\rho ({\bf r}_{1},{\bf r}%
_{3})\,\left( I_{1324}+I_{2424}\right) \right] \right\} $ \\ 
\  & \  \\ 
\  & $-\frac{c}{2}\left\{ 4\left[ \rho ({\bf r}_{1},{\bf r}_{3})\,\rho ({\bf %
r}_{2},{\bf r}_{4})-\delta _{\tau _{1}\tau _{2}}\delta _{s_{1}s_{2}}\,\rho (%
{\bf r}_{1},{\bf r}_{4})\,\rho ({\bf r}_{2},{\bf r}_{3})\right] \left(
I_{1}+I_{2}+I_{3}+I_{4}\right) \right. $ \\ 
\  & \  \\ 
\  & $-\rho ({\bf r}_{1},{\bf r}_{3})\left(
I_{124}+I_{224}+I_{324}+I_{424}\right) -\rho ({\bf r}_{2},{\bf r}_{4})\left(
I_{113}+I_{213}+I_{331}+I_{413}\right) $ \\ 
\  & \  \\ 
\  & $+\left. \delta _{\tau _{1}\tau _{2}}\delta _{s_{1}s_{2}}\,\left[ \rho (%
{\bf r}_{2},{\bf r}_{3})\left( I_{114}+I_{214}+I_{314}+I_{414}\right) -\rho (%
{\bf r}_{1},{\bf r}_{4})\left( I_{123}+I_{223}+I_{323}+I_{423}\right)
\right] \right\} ,$%
\end{tabular}
\end{equation}
while the last term $\rho _{C}^{(2)}$ reads:

\begin{equation}
\begin{tabular}{ll}
$\rho _{C}^{(2)}=$ & $c{\cal M}({\bf r}_{1},s_{1},\tau _{1};{\bf r}%
_{2},s_{2},\tau _{2};{\bf r}_{3},s_{1},\tau _{1};{\bf r}_{4},s_{2},\tau
_{2};\ y)-\frac{1}{2}\ c^{2}{\cal M}({\bf r}_{1},s_{1},\tau _{1};{\bf r}%
_{2},s_{2},\tau _{2};{\bf r}_{3},s_{1},\tau _{1};{\bf r}_{4},s_{2},\tau
_{2};2y)\ \ ,$%
\end{tabular}
\end{equation}

\smallskip 
\begin{equation}
\begin{tabular}{ll}
${\cal M}$ & $=4\left[ \rho ({\bf r}_{1},{\bf r}_{3})\ Y_{24}-\rho ({\bf r}%
_{2},{\bf r}_{4})\ Y_{13}\right] -Y_{1324}+\rho ({\bf r}_{1},{\bf r}_{3})\
Z_{42}-\rho ({\bf r}_{2},{\bf r}_{4})\ Z_{13}$ \\ 
\  & \  \\ 
\  & $-\delta _{\tau _{1}\tau _{2}}\delta _{s_{1}s_{2}}\left\{ 4\left[ \rho (%
{\bf r}_{1},{\bf r}_{4})Y_{23}-\rho ({\bf r}_{2},{\bf r}_{3})Y_{14}\right]
-Y_{1423}-\rho ({\bf r}_{2},{\bf r}_{3})Z_{14}-\rho ({\bf r}_{1},{\bf r}%
_{4})Z_{23}\right\} .$%
\end{tabular}
\end{equation}
where the following products of exponential and polynomial factors depending
on the scaled coordinates ${\bf x}_{i}=\alpha {\bf r}_{i}$ and the
dimensionless parameter $y=\beta ^{2}/\alpha ^{2}$ are introduced:

\begin{equation}
\begin{tabular}{rl}
$I_{1}$ & $={\it exp}[{-\frac{yx_{1}^{2}}{1+y}}]\,P_{I_{1}}({\bf x}_{1})$ \\ 
\  & \  \\ 
$I_{12}$ & $={\it exp}[{-{\frac{y\,\left( {x_{1}}^{2}+\ {x_{2}}^{2}+\ 2y\,\ {%
({\bf x}_{1}-{\bf x}_{2})^{2}}\right) }{{1+2\,y}}}}]\,P_{I_{2}}({\bf x}_{1},%
{\bf x}_{2})$ \\ 
\  & \  \\ 
$I_{123}$ & $={\frac{\alpha ^{3}}{\pi ^{3/2}}}\,\ {\it exp}[{-\frac{%
x_{2}^{2}\ +\ x_{3}^{2}\ +\ y\ (2x_{1}^{2}+x_{2}^{2}+x_{3}^{2})}{2(1+y)}}%
]\,P_{I_{3}}({\bf x}_{1},{\bf x}_{2},{\bf x}_{3})$ \\ 
\  & \  \\ 
$I_{1234}$ & $=\frac{\alpha ^{3}}{\pi ^{3/2}}\ {\it \,exp}[-\frac{%
x_{3}{}^{2}\ +\ x_{4}{}^{2}\ +\ 2\,y\,\ (x_{1}{}^{2}+x_{2}{}^{2})\ +\
2\,y^{2}\,({\bf x}_{1}-{\bf x}_{2})^{2}}{2\,(1+2\,y)}]\,P_{I_{4}}({\bf x}%
_{1},{\bf x}_{2},{\bf x}_{3},{\bf x}_{4})$%
\end{tabular}
\label{i1}
\end{equation}

and

\begin{equation}
\begin{tabular}{rl}
$Y_{12}$ & $={\frac{\alpha ^{3}}{\pi ^{3/2}}}\,\ {\it exp}[{-\frac{%
x_{1}^{2}\ +\ x_{2}^{2}}{2}}]\,P_{Y_{2}}({\bf x}_{1},{\bf x}_{2};\ y)$ \\ 
\  & \  \\ 
$Z_{12}$ & $={\frac{\alpha ^{3}}{\pi ^{3/2}}\ }\,{\it exp}[{-\frac{%
x_{1}^{2}\ +\ x_{2}^{2}}{2}}]\,P_{Z_{2}}({\bf x}_{1},{\bf x}_{2};\ y)$ \\ 
\  & \  \\ 
$Y_{1234}$ & $={\frac{\alpha ^{6}}{\pi ^{3}}\ }\,{\it exp}[{-\frac{%
x_{1}^{2}\ +\ x_{2}^{2}\ +\ x_{3}^{2}\ +\ x_{4}^{2}}{2}}]\,P_{Y_{3}}({\bf x}%
_{1},{\bf x}_{2},{\bf x}_{3},{\bf x}_{4};\ y)$%
\end{tabular}
\label{y3p}
\end{equation}

\smallskip

The two-body density matrix represented by\ eqs.(\ref{rabc}-\ref{y3p}) does
not depend on the particular choice of the nucleus. All nuclear structure
information is reflected by the polynomials $\left\{
P_{I},P_{Y},P_{Z}\right\} $ entering eqs.(\ref{i1},\ref{y3p}). Their
explicit form is given in the Appendix for $^{4}He$ and $^{16}O$ nuclei. As
a result we end up with closed analytical algebraic expressions for the
two-body density matrix involving three independent parameters, the harmonic
oscillator length $\alpha $ and the correlations parameters $c$ and $\beta $%
. Due to its gaussian-polynomial structure one can easily derive the Fourier
transforms defining the associated two-body density matrix in the momentum
space.

\section{Local density and momentum pair distributions}

Having derived the two-body density matrix in the coordinate space one is
able to calculate all two--body nuclear characteristics in close analytical
form. Changing the coordinates of two particles ${\bf r}_{1}$ and ${\bf r}_{2%
\text{ }}$in eqs.(\ref{rabc}-\ref{y3p}) to the center-of-mass ${\bf R=}%
\left( {\bf r}_{1}+{\bf r}_{2}\right) {\bf /}2$ and relative ${\bf s=}\left( 
{\bf r}_{1}{\bf -r}_{2}\right) $ coordinates, one obtains the center-of-mass 
$\rho ^{(2)}({\bf R})$ , eq.(\ref{denrelR}), and relative ${{\rho ^{(2)}(%
{\bf s})}}$ , eq.(\ref{denrelr}), pair local distributions. Using the
dimensionless variables $R=\alpha |{\bf R|}$ and $s=\alpha |{\bf s|}$ they
have the following form:

\begin{equation}
\begin{tabular}{ll}
$\rho ^{(2)}(R)$ & $={\cal A}{\it exp}\left[ {-}\left( \sqrt{2}R\right)
^{2}\right] +{\cal B}{{\it exp}\left[ {-\frac{\,2(1+2\,y)}{2+3\,y}}\left( 
\sqrt{2}R\right) ^{2}\right] +}{\cal C}{{\it exp}\left[ -{\frac{\,1+4\,y}{%
1+3\,y}}\left( \sqrt{2}R\right) ^{2}\right] \ ,}$%
\end{tabular}
\label{r2R}
\end{equation}

\begin{equation}
\begin{tabular}{ll}
$\rho ^{(2)}{(}s{)}$ & $={\cal D}{\it exp}\left[ -\left( s/\sqrt{2}\right)
^{2}\right] +{\cal E}\,{\it exp}\left[ {-\frac{2(1+2\,y)}{2+3\,y}}\left( s/%
\sqrt{2}\right) ^{2}\right] {+}{\cal F}{\it exp}\left[ {-\frac{1+4\,y}{1+3\,y%
}}\left( s/\sqrt{2}\right) ^{2}\right] $ \\ 
&  \\ 
& $+{\cal G}{\it exp}\left[ {-}(1+4\,y)\left( s/\sqrt{2}\right) ^{2}\right] -%
{\cal H}\,{\it exp}\left[ {-}(1+2\,y)\left( s/\sqrt{2}\right) ^{2}\right] \ ,
$%
\end{tabular}
\label{r2s}
\end{equation}
where 
\begin{equation}
\begin{tabular}{lll}
${\cal A}={\frac{\alpha ^{3}}{\pi {^{3/2}}}}[{{\eta }_{1}(R)+2\ c\,{\eta }%
_{2}(R,y)-\,c^{2}\,{\eta }_{2}(R,2\,y)],}$ & ${\cal B}={\frac{\alpha ^{3}}{%
\pi {^{3/2}}}c\,{\eta _{3}(R),}}$ & ${\cal C}{=\frac{\alpha ^{3}}{\pi {^{3/2}%
}}c^{2}\,{\eta _{4}(R),}}$%
\end{tabular}
\end{equation}

\begin{equation}
\begin{tabular}{lll}
${\cal D}={\frac{\alpha ^{3}}{\pi ^{3/2}}\,}\left[ {\mu _{1}(s)+2\,c\,\mu
_{2}(s;y)-{c^{2}}\,\mu _{2}(s;2y)}\right] ,$ & ${\cal E}=-{\frac{\alpha ^{3}%
}{\pi ^{3/2}}}2\,c\,\mu _{3}(s;y),$ & ${\cal F}={\frac{\alpha ^{3}}{\pi
^{3/2}}}c^{2}\,\,\mu _{3}(s;2y)$\newline
$,$ \\ 
\  & \  &  \\ 
\  & ${\cal G}={\frac{\alpha ^{3}}{\pi ^{3/2}}}c^{2}\mu _{1}(s),$ & ${\cal H}%
=-{\frac{\alpha ^{3}}{\pi ^{3/2}}}2\,c\mu _{1}(s).$%
\end{tabular}
\end{equation}

The center-of-mass $n^{(2)}({\bf K})$ , eq.(\ref{denrelK}), and relative $%
n^{(2)}({\bf k})$ , eq.(\ref{denrelk}), pair momentum distributions follow
after performing Fourier transform of the center-of-mass ${\rho }^{(2)}{(%
{\bf R},}{\bf R}^{\prime }{)}$ and relative ${\rho }^{(2)}{({\bf s},}{\bf s}%
^{\prime }{)}$ pair\ density matrices, respectively. Again, introducing $K=$ 
$|{\bf K|}$ ${\bf /}\alpha $ and $k=|$ ${\bf k|\ /\alpha }$ one obtains:

\begin{equation}
\begin{tabular}{ll}
${{n^{(2)}({\bf K})}}$ & $={\cal A}^{\prime }{\it exp}\left[ -\left( \frac{K%
}{\sqrt{2}}\right) ^{2}\right] +{\cal B}^{\prime }{\it exp}\left[ -{\frac{%
1+2\,y}{\,2+5\,y}}\left( \frac{K}{\sqrt{2}}\right) ^{2}\right] {+}{\cal C}%
^{\prime }{\it exp}\left[ -{\frac{1}{\,(1+y)}}\left( \frac{K}{\sqrt{2}}%
\right) ^{2}\right] ,$%
\end{tabular}
\label{nK}
\end{equation}

\begin{equation}
\begin{tabular}{ll}
${{n^{(2)}({\bf k})}}$ & $={\cal D}^{\prime }{\it exp}\left[ -\left( \sqrt{2}%
k\right) ^{2}\right] +{\cal E}^{\prime }\,{\it exp}\left[ -{\frac{2\left(
1+2\,y\right) }{2+5\,y}}\left( \sqrt{2}k\right) ^{2}\right] {+}{\cal F}%
^{\prime }{\it exp}\left[ -{{{\frac{\,{1}}{1+y}}}}\left( \sqrt{2}k\right)
^{2}\right] $ \\ 
\  & \  \\ 
\  & $+{\cal G}^{\prime }{\it exp}\left[ -{\frac{\,{1}}{1+4\,y}}\left( \sqrt{%
2}k\right) ^{2}\right] -{\cal H}^{\prime }{\it exp}\left[ -{\frac{\,1+2\,y}{%
1+4\,y}}\left( \sqrt{2}k\right) ^{2}\right] ,$%
\end{tabular}
\label{nk}
\end{equation}
where

\begin{equation}
\begin{tabular}{lll}
${\cal A}^{\prime }={\frac{\sqrt{2}}{2\alpha ^{3}\pi ^{3/2}}}\left( {\gamma }%
_{1}(K)+2c\,{\gamma }_{2}(K,y)-c^{2}\,{\gamma }_{2}(K,2\,y)\right) {,}$ & $%
{\cal B}^{\prime }={\frac{\sqrt{2}}{2\alpha ^{3}\pi ^{3/2}}}c\,{\gamma }%
_{3}(K)\,{,}$ & ${\cal C}^{\prime }{={\frac{\sqrt{2}}{2\alpha ^{3}\pi ^{3/2}}%
}c^{2}\,{\gamma }_{4}(K){,}}$%
\end{tabular}
\end{equation}

\begin{equation}
\begin{tabular}{lll}
${\cal D}^{\prime }={\frac{\sqrt{2}}{2\alpha {^{3}}\,{\pi }^{3/2}}}\left(
\theta _{1}(k)+2c\,\theta _{2}(k,y)-\,c^{2}\,{\theta }_{2}(k,2\,y)\right) ,$
& ${\cal E}^{\prime }={\frac{\sqrt{2}}{2\alpha {^{3}}\,{\pi }^{3/2}}}$%
\newline
$c{\theta }_{4}(k),$ & ${\cal F}^{\prime }={\frac{\sqrt{2}}{2\alpha {^{3}}\,{%
\pi }^{3/2}}}c^{2}\,{\theta }_{6}\,(k),$ \\ 
\  & \  & \  \\ 
\  & ${\cal G}^{\prime }={\frac{\sqrt{2}}{2\alpha {^{3}}\,{\pi }^{3/2}}}%
c^{2}\,{\theta }_{5}\,(k),$ & ${\cal H}^{\prime }={\frac{\sqrt{2}}{2\alpha {%
^{3}}\,{\pi }^{3/2}}}c{\theta }_{3}(k)\,.$%
\end{tabular}
\end{equation}

In the above expressions the exponential dependence is explicit while the
expressions for the polynomial amplitudes $\left\{ \eta _{i}\right\} $, $%
\left\{ \mu _{i}\right\} $, $\left\{ \gamma _{i}\right\} $ and $\left\{
\theta _{i}\right\} $ are presented in the Appendix for $^{4}He$ and $^{16}O$
nuclei.

For completeness, we are giving also the local nuclear momentum distribution 
$n(k)$, eq.(\ref{nucmom}), which is associated with the one-body density
matrix. Its closed analytical form within the present model has already been
derived in \cite{r9}:

\begin{equation}
n({\bf k})={\cal A}^{\prime \prime }\exp \left[ -k^{2}\right] +{\cal B}%
^{\prime \prime }{\exp }\left[ -{\frac{1}{{1+2\,y}}}k^{2}\right] -{\cal C}%
^{\prime \prime }{\exp }\left[ -{\frac{{\,1+2\,y}}{{1+3\,y}}}k^{2}\right] .
\label{label3}
\end{equation}
and the polynomial expressions for the amplitudes ${\cal A}^{\prime \prime },%
{\cal B}^{\prime \prime }{,}{\cal C}^{\prime \prime }$ can be found in \cite
{r9}. Considering the gaussian factors one can realize a factor $\sqrt{2}$
which scales the momenta entering the three momentum distributions ${{%
n^{(2)}({\bf K}),\;n({\bf k})}}$ and ${{n^{(2)}({\bf k})}}$.

\section{Results and discussion}

Next step in our study is to define the parameters of the problem, the
oscillator parameter $\alpha $ and the parameters $\beta $ and $c$ related
to  the healing  distance and the strength of the SRC, respectively. The
rigorous procedure would be to apply the variational approach based on the
one- and two-body density matrices defined so far by minimizing the total
energy of the system with a realistic $NN-$interaction with respect to the
parameters $\alpha $, $\beta $ and $c$. As in our previous papers\cite{r9},
however, we prefer to obtain the values of $\alpha $ and $\beta $
phenomenologically by fitting the experimental elastic formfactor data using
the analytical expression for the elastic formfactor following within the
present model from eq.(\ref{ff}). The values of the parameter $c$ for $^{4}He
$ and $^{16}O$ are determined under the additional condition the relative
pair density distribution ${\rho }^{\left( 2\right) }{({\bf s})}$, eq.(\ref
{r2s}), to reproduce at $s=0$ the associated value obtained within the
Variational Monte-Carlo approach \cite{car} and \cite{pwp}, respectively,
while we simply put $c=1$ for $^{40}Ca$ since there are no variational
calculations for $^{40}Ca$ . Thus, in the present numerical calculations we
are using the following values of the parameters: 
\begin{equation}
\begin{tabular}{llll}
$\alpha =0.82\;fm^{-1},$ & $\beta =1.23\;fm^{-1},$ & $c=0.76$ & for $^{4}He\
,$ \\ 
$\alpha =0.61\;fm^{-1},$ & $\beta =1.30\;fm^{-1},$ & $c=0.77$ & for $^{16}O\
,$ \\ 
$\alpha =0.52\;fm^{-1},$ & $\beta =1.21\;fm^{-1},$ & $c=1.00$ & for $%
^{40}Ca\ $.
\end{tabular}
\end{equation}

At the beginning we are comparing in Fig. 1 some results from our crude but
analytical model (the solid lines) with the results (dashed lines) emerging
from the orders of magnitude more complicated Variational Monte-Carlo
calculations \cite{car}, \cite{pwp}. The top panel of Fig. 1 clearly
demonstrates that our simple gaussian form for the correlation factor $f(r)$
, eq.(\ref{fcorr}), should be considered only as a first approximation to
the actual shape of the realistic central correlation function.
Nevertheless, the analytical results for the relative density pair
distributions in $^{4}He$ and $^{16}O$ ( two bottom panels in Fig. 1) show
quite acceptable quantitative agreement with the Variational Monte-Carlo
results. Despite of the simplicity of the model and the LOA used, the
characteristic behavior of the realistic pair density is obviously
reproduced. The small values of the relative pair distributions at $s=0$
indicate the presence of significant SRC which in the Variational
Monte-Carlo calculations are due to the repulsive core of the two-body
interaction.

In Fig. 2 a comparison is made between the correlated ($c\neq 0$ , solid
curves) and the uncorrelated ($c=0$, dashed curves) results for the relative
pair density distribution ${\rho }^{\left( 2\right) }{({\bf s})}$ , eq.(\ref
{r2s}), for $^{4}He$, $^{16}O$ and $^{40}Ca$. It is seen that due to the SRC
the shape of the distributions changes significantly. The SRC lead to a deep
hole in the correlated distributions near $s=0$, while the uncorrelated
distributions saturate at small distances to values which are significantly
different from zero. Our calculations have shown that the SRC do not affect
significantly the center-of-mass pair local distributions in the coordinate
space.

The above examples show that despite of the simplicity of the model it is
able to incorporate the SRC into the two-body quantities of interest.
Therefore it may be considered as a useful starting point in analyzing
two-particle emission experiments.

From $(e,e^{\prime }2N)$ experiments one can expect important information
about the effects of SRC on the pair local distributions in the momentum
space. In Fig. 3 we present the center-of-mass (solid lines) $n^{(2)}(K)$,
eq.(\ref{nK}), and relative (dot-dashed lines) $n^{(2)}(k)$, eq.(\ref{nk}),
pair momentum distributions for $^{4}He$, $^{16}O$ and $^{40}Ca$ . A
comparison is also made between the correlated $(c\neq 0)$ and uncorrelated $%
(c=0)$ results. For completeness, by dashed lines in Fig. 3, we also show
the results for the local nuclear momentum distribution $n(k)$, eq.(\ref
{label3}), which is associated with the one-body density matrix.

From Fig. 3 is clearly seen that due to the SRC high momentum tails develop
at large values of the momenta and this behavior is typical for all three
kinds of momentum distributions considered. Comparing with the uncorrelated
results one can see that the correlation effects start to dominate first for
the relative pair momentum distribution $n^{(2)}(k)$. Then, at larger values
of the momenta, the center-of-mass distribution $n^{(2)}(K)$ also develops a
high momentum tail. The local nuclear momentum distribution $n(k)$ takes an
intermediate position between both distributions $n^{(2)}(k)$ and $n^{(2)}(K)
$. In the case of $^{4}He$ for example the SRC become important for momenta
larger than $1.4,\;2.1,\;2.7$ $fm^{-1}$, for the momentum distributions $%
n^{(2)}(k),\;n(k)$ and $n^{(2)}(K)$, respectively. Similar values are valid
also for nuclei $^{16}O$ and $^{40}Ca$. The reason for such behavior is
obviously due to the scaling factor of $\sqrt{2}$ which has been observed in
the analytical expressions for $n^{(2)}(K)$, eq.(\ref{nK}), $n(k)$, eq.(\ref
{nK}) and $n^{(2)}(k)$, eq.(\ref{nk}). Also, at large momenta both the
correlated and uncorrelated momentum distributions obviously satisfy the
inequality: 
\[
n^{(2)}(k)<\;n(k)<n^{(2)}(K)\;.
\]

Of course, the comparison of the results given in Fig. 3 has to be done
keeping in mind the different meaning of the arguments of the three types of
momentum distributions considered..

Plotting together each of the nucleon momentum distributions $%
n^{(2)}(k),\;n(k)$ and $n^{(2)}(K)$ for all nuclei $^{4}He$, $^{16}O$ and $%
^{40}Ca$, as it is done in Fig. 4, one can detect the interesting fact that
in the high momentum region these distributions are almost universal in the
sense that they do not significantly depend on the mass number $A$. This
universal behavior of the local nucleon momentum distribution $n(k)$ has
been observed earlier (see e.g. \cite{pwp,r9,r13}). Obviously similar
tendency exists also for the relative and center-of-mass pair momentum
distributions.

\section{Conclusions}

In this paper we have derived closed analytical expressions for the two-body
density matrix and associated two-body nuclear characteristics within the
LOA to the Jastrow correlation method for the closed $s-d$ shell nuclei $%
^{4}He,^{16}O$ and $^{40}Ca$ under two simplifying assumptions:
harmonic-oscillator single-particle wave functions entering the Slater
determinant and a state-independent gaussian-like correlation function. The
comparison with more realistic results emerging from Variational Monte-Carlo
calculations has shown the usefulness of the expressions derived. They can
be applied as a starting tool in analyzing two-particle emission experiments
where important information about the effects of SRC on the two-body nuclear
characteristics is expected. In particular, it has been shown that the SRC
effects start to dominate in the high-momentum region of the relative $%
n^{(2)}(k),$ local $n(k)$ and center-of-mass $n^{(2)}(K)$ momentum
distributions at different momentum values which are proportional to a
factor of $\sqrt{2}$ and are almost independent on the nucleus considered.
An universal asymptotic behavior of the relative and center-of-mass pair
momentum distributions (normalized to unity) is indicated similar to the
well known high-momentum tail of the nucleon momentum distribution $n(k)$.

As a first approximation, the simplicity of the results can help us to
concentrate our attention towards the complicated questions arising from the
involved physical interpretation and the mechanism of the two--particle
emission processes in nuclei.

{\bf Acknowledgments}

One of the authors (S.S.D) is grateful to Dr. P.E. Hodgson and the Royal
Society for the warm hospitality and financial support at the Astrophysics
and Nuclear Physics Laboratory of the Oxford University. This work is
supported in part by the Contract $\Phi -809$ with the Bulgarian National
Science Foundation.

\section{Appendix}

\subsection{Polynomials entering the two-body density matrix}

The expressions for  $\left\{ P\right\} $ in eqs.(\ref{i1},\ref{y3p}) which
determine the two-body density matrix are:

\subsubsection{Nucleus $^{4}He$}

\[
\begin{tabular}{l}
${P_{I_{1}}=P_{I_{2}}={(1+2y)}^{-3/2}}$ \\ 
\  \\ 
${P_{I_{3}}=P_{I_{4}}={(1+y)}^{-3/2}}$ \\ 
\  \\ 
${P_{Y_{2}}(y)=P_{Z_{2}}(y)=P_{Y_{3}}(y)={(1+2y)}^{-3/2}}$%
\end{tabular}
\]

\subsubsection{Nucleus $^{16}O$}

\[
\ 
\begin{tabular}{rl}
${P_{I_{1}}({\bf x})}$\  & ${={\frac{1}{(1+y)^{7/2}}}\left[ 4+5\,y+{y^{2}~}%
\left( 1+2\,{{x}^{2}}\right) \right] }$ \\ 
&  \\ 
${P_{I_{2}}({\bf x}_{1}{,{\bf x}}_{2})}$\  & ${={\frac{2}{{\ ({1+2\,y)}^{7/2}%
}}}{\left[ 2+5\,y+{y^{2}}~(2+{x_{1}}^{2}+{x_{2}}^{2}+2x_{12})\right] }}$\ 
\\ 
&  \\ 
$P_{I_{3}}({\bf x}_{1},{\bf x}_{2},{\bf x}_{3})$\  & $={{\frac{1}{(1+y)^{7/2}%
}}}\left[ {1+2\,{{x_{23}}}+2\,y~\left( 1+\,{{x_{12}}}\,+\,{{x_{13}}}\,+\,{{%
x_{23}}}\,\right) +y^{2}~(1+2\,{{x_{12}}}\,+2\,{x_{13}}\,+4\,{x_{12~}x_{13})}%
}\right] $ \\ 
& ${\,}$ \\ 
$P_{I_{4}}({\bf x}_{1},{\bf x}_{2},{\bf x}_{3},{\bf x}_{4})$\ \  & $={{\frac{%
1}{({{1+2\,y)}^{{7/2}}}}}\left[ 1+2\,{{x_{34}}}+2\,y~(~2\,+{{x_{13}}}\,+{{%
x_{14}}}\,+{{x_{23}}}\,+{{x_{24}}}\,+2\,{{x_{34}}}\,)\right. }$ \\ 
&  \\ 
\  & ${\left. +\ \,{4\,{y^{2}~(}+{{x_{13}}}\,+{{x_{14}}}+{{x_{23}}}+{{x_{24}}%
}\,+{{x_{23}}}\,+{x_{13}}\,{x_{14}}\,+{x_{14}}}\,{{x_{23}}}\,+{{x_{13~}}}{%
x_{24}}\,{{x_{24})}}\right] }$%
\end{tabular}
\]

\[
\begin{tabular}{rl}
$P_{Y_{2}}({\bf x}_{1},{\bf x}_{2};\ y)$ & $={{\frac{1}{(1+2\,y)^{7/2}}}%
\left[ 4+8\,{{x_{12}}}+y\,~\left( 13+18\,{{x_{12}}}\right) +{y^{2}}\,\left(
10+14\,{{x_{12}}}\right) \right] }$ \\ 
\  &  \\ 
$P_{Z_{2}}({\bf x}_{1},{\bf x}_{2};\ y)$ & $={{\frac{1}{(1+2\,y)^{7/2}}}%
\left[ 1+2\,{{x_{12}}}+y\,~\left( 7+6\,{{x_{12}}}\right) +{y^{2}}~\left(
10+14\,{{x_{12}}}\right) \right] }$ \\ 
\  &  \\ 
$P_{Y_{3}}({\bf x}_{1},{\bf x}_{2},{\bf x}_{3},{\bf x}_{4};\ y)$ & $={{\frac{%
1}{(1+2\,y)^{7/2}}}~\left[ 1+2\,{{x_{12}}}+2\,{{x_{34}}}+4\,{{x_{12}}}\,{{%
x_{34}+}}\right. }$ \\ 
\  &  \\ 
\  & $2\,y~(2+3\,{{x_{12}}}+{{{x_{13}}}+{{x_{14}}}+{{x_{23}}}+{{x_{24}}}+3\,{%
{x_{34}}}+4\,{{x_{12}~x_{34}}})}$ \\ 
\  &  \\ 
\  & ${\left. +{{4\,y^{2}~(}1+{{x_{12}}}+{{x_{13}}}+{{x_{14}}}+{{x_{23}}}+{{%
x_{24}}}+{{x_{34}}}+{x_{14}~x_{23}\ }}~+{{x_{13~}x_{24}}}+{{x_{12}~x_{34}}}%
)~\right] }$%
\end{tabular}
\]

\subsubsection{Nucleus $^{40}Ca$}

The expressions for $^{40}Ca$ have similar structure but are up to 4-th
order with respect to $x_{ij}$, thus being too long to be presented here.
They can be obtained in the form of user friendly files for Mathematica 3.0
upon request.

\subsection{Polynomials entering local density pair distributions}

The expressions for $\left\{ \eta _{i}\right\} $, $\left\{ \mu _{i}\right\} $%
, $\left\{ \gamma _{i}\right\} $ and $\left\{ \theta _{i}\right\} $, which
determine the two-body density and momentum distributions eqs.(\ref{r2R}-\ref
{nk}) are:

\subsubsection{Nucleus $^{4}He$}

\[
\begin{tabular}{llll}
$\eta _{1}{(R)}=12\sqrt{2}$ & $\eta _{2}{(R;\ y)}=-\frac{4\sqrt{2}}{(1+2\
y)^{3/2}}$ & $\eta _{3}{(R)}=\frac{32}{(2+3\ y)^{3/2}}$ & ${{\eta _{4}(R)=-\ 
}}\frac{4\sqrt{2}}{(1+3\ y)^{3/2}}$ \\ 
\  &  &  &  \\ 
$\mu _{1}(s)=\frac{3}{\sqrt{2}}$ & $\mu _{2}(s;\ y)=\frac{15}{\sqrt{2}(1+2\
\,y)^{3/2}}$ & $\mu _{3}(s;\ y)={\frac{24}{(2+3\,y)^{3/2}}}$ &  \\ 
&  &  &  \\ 
$\gamma _{1}(K)=\frac{1}{2}$ & $\gamma _{2}(K,y)=\frac{2}{(1+2\,y)^{3/2}}$ & 
$\gamma _{3}(K)=-\ {\frac{8\,\sqrt{2}}{{(2+5\,y)}^{3/2}}}$ & $\gamma _{4}(K)=%
{\frac{2}{(1+y)^{3/2}(1+4\,y)^{3/2}}}$ \\ 
\  &  &  &  \\ 
$\theta _{1}(k)=4$ & $\theta _{2}(k;\ y)=\frac{20}{(1+y)^{3/2}}$ & $\theta
_{3}(k)=\frac{8}{(1+4\ y)^{3/2}}$ &  \\ 
\  &  &  &  \\ 
$\theta _{4}(k)=-\frac{64\sqrt{2}}{(2+5\ y)^{3/2}}$ & $\theta _{5}(k)=\frac{4%
}{(1+4\ y)^{3}}$ & $\theta _{6}(k)=\frac{16}{(1+\ y)^{3/2}\ (1+4\ y)^{3/2}}$
& 
\end{tabular}
\]

\subsubsection{Nucleus $^{16}O$}

\[
\begin{tabular}{rl}
$\eta _{1}{(R)}$ & $=2\,\sqrt{2}\,\left( 117+104\,{R^{2}}+48\,{R^{4}}\right) 
$ \\ 
\  &  \\ 
$\eta _{2}{(R;\ y)}$ & $={\frac{\sqrt{2}}{(1+2\,y)^{7/2}}}\,\left[ 48\,{R^{4}%
}\,(28+60\,y+35\,{y^{2}})\right. $ \\ 
\  &  \\ 
\  & $+\left. \ 8\,{R^{2}}\,(364+848\,y+555\,{y^{2}})+3\,(1092+2780\,y+1837\,%
{y^{2}})\right] $ \\ 
\  &  \\ 
$\eta _{3}{(R)}$ & $=-\ \frac{{128}}{({2+3\,y)}^{15/2}}\ \left[ 1536\,{R^{6}}%
\,{y^{2}}\,{{(1+3\,y+2\,{y^{2}})}^{2}}\right. $ \\ 
\  &  \\ 
\  & $+\ 192\,{R^{4}}\,(28+224\,y+749\,{y^{2}}+1352\,{y^{3}}+1419\,{y^{4}}%
+843\,{y^{5}}+225\,{y^{6}})$ \\ 
\  &  \\ 
\  & $+\ 8\,{R^{2}}\,{{(2+3\,y)}^{2}}\,(364+1762\,y+3489\,{y^{2}}+3105\,{%
y^{3}}+990\,{y^{4}})$ \\ 
\  &  \\ 
\  & $+\ \left. 9\,{{(2+3\,y)}^{3}}\,(182+563\,y+588\,{y^{2}}+212\,{y^{3}}%
)\right] $ \\ 
\  &  \\ 
$\eta _{4}(R)$ & $=\frac{4\sqrt{2}}{({1+3\,y)}^{15/2}}\ \left[ 384\,{R^{6}}\,%
{y^{2}}\,{{(1+6\,y+8\,{y^{2}})}^{2}}\right. $ \\ 
\  &  \\ 
\  & $+\ 48\,{R^{4}}\,(7+112\,y+749\,{y^{2}}+2704\,{y^{3}}+5676\,{y^{4}}%
+6744\,{y^{5}}+3600\,{y^{6}})$ \\ 
\  &  \\ 
\  & $+\ 8\,{{R^{2}\,(1+3\,y)}^{2}}\,(91+881\,y+3489\,{y^{2}}+6210\,{y^{3}}%
+3960\,{y^{4}})$ \\ 
\  &  \\ 
\  & $+\ \left. 9\,{{(1+3\,y)}^{3}}\,(91+563\,y+1176\,{y^{2}}+848\,{y^{3}}%
)\right] $%
\end{tabular}
\]
\[
\begin{tabular}{rl}
$\mu _{1}(s)$ & $={\frac{1}{4\ \sqrt{2}}\ (}93+34\,{s^{2}}+3\,{s^{4})}$ \\ 
\  &  \\ 
$\mu _{2}(s;\ y)$ & $=\frac{1}{4\ \sqrt{2}(1+2\,y)^{7/2}}\ \left[
2697+6540\,y+4047\,{y^{2}}\right. $ \\ 
\  &  \\ 
& $+\left. 2\,{s^{2}}\,(493+1214\,y+831\,{y^{2}})+3\,{s^{4}}\,(29+64\,y+39\,{%
y^{2}})\right] $ \\ 
\  &  \\ 
$\mu _{3}(s;\ y)$ & $={\frac{8}{(2+3\,y)^{15/2}}\ }\left[ 3\,{{(2+3\,y)}^{3}}%
\,(434+1245\,y+1158\,{y^{2}}+348\,{y^{3}})\right. $ \\ 
\  &  \\ 
& $+\ 2\,{s^{2}}\,{{(2+3\,y)}^{2}}\,(476+2370\,y+4713\,{y^{2}}+4215\,{y^{3}}%
+1386\,{y^{4}})$ \\ 
\  &  \\ 
& $+\ 12\,{s^{4}}\,(28+232\,y+813\,{y^{2}}+1554\,{y^{3}}+1733\,{y^{4}}+1083\,%
{y^{5}}+297\,{y^{6}})$ \\ 
\  &  \\ 
& $+\ \left. 24\,{s^{6}}\,{y^{2}}\,{{(1+3\,y+2\,{y^{2}})}^{2}}\right] $%
\end{tabular}
\]
\[
\begin{tabular}{rl}
$\gamma _{1}(K)$ & $=\frac{1}{480}\ (117+26\,{K^{2}}+3\,{K^{4})}$ \\ 
\  &  \\ 
${\gamma _{2}(K;\ y)}$ & $={{\frac{1}{480\,{(1+2\,y)}^{7/2}}}\left[ 3\,{K^{4}%
}\,\left( 28+60\,y+35\,{y^{2}}\right) \right. }$ \\ 
\  &  \\ 
\  & $+\left. 2\,{K^{2}}\,\left( 364+872\,y+603\,{y^{2}}\right) +3\,\left(
1092+2732\,y+1741\,{y^{2}}\right) \right] $ \\ 
\  &  \\ 
${\gamma _{3}(K)}$ & $={{\frac{2\sqrt{2}}{15\,{(2+5\,y)}^{15/2}}}\,\left[
24\,{K^{6}}\,{y^{2}}\,{{(1+2\,y)}^{2}}\,(1+4\,y+3\,{y^{2}})\right. }$ \\ 
\  &  \\ 
\  & $-\ 12\,{K^{4}}\,(28+344\,y+1705\,{y^{2}}+4340\,{y^{3}}+5933\,{y^{4}}%
+4087\,{y^{5}}+1105\,{y^{6}})$ \\ 
\  &  \\ 
\  & $-\ 2\,{K^{2}}\,{{(2+5\,y)}^{2}}\,(364+2870\,y+8263\,{y^{2}}+10185\,{%
y^{3}}+4680\,{y^{4}})$ \\ 
\  &  \\ 
\  & $-\ \left. 3\,{{(2+5\,y)}^{3}}\,(546+3251\,y+6160\,{y^{2}}+3770\,{y^{3}}%
)\right] $ \\ 
\  &  \\ 
${\gamma _{4}(K)}$ & $={{\frac{1}{120\,{(1+y)}^{11/2}\,{(1+4\,y)}^{7/2}}}%
\left[ 3\,{K^{4}}\,(7+46\,y+72\,{y^{2}})\right. }$ \\ 
\  &  \\ 
\  & $+\ 2\,{K^{2}}\,(91+707\,y+2250\,{y^{2}}+3238\,{y^{3}}+1604\,{y^{4}})$
\\ 
\  &  \\ 
\  & $+\ \left. 3\,{{(1+y)}^{2}}\,(273+1886\,y+5060\,{y^{2}}+6424\,{y^{3}}%
+3472\,{y^{4}})\right] $%
\end{tabular}
\]

\[
\begin{tabular}{rl}
$\theta _{1}(k)$ & $=\frac{1}{60}\ {(}117+104\,{k^{2}}+48\,{k^{4})}$ \\ 
\  &  \\ 
$\theta _{2}(k;\ y)$ & $={\frac{1}{60\,{{(1+2\,y)}^{7/2}}}\ }\left[ 48\,{%
k^{4}}\,(29+64\,y+39\,{y^{2}})\right. $ \\ 
\  &  \\ 
\  & $\left. +\ 8\,{k^{2}}\,(493+1202\,y+807\,{y^{2}})+3\,(899+2204\,y+1397\,%
{y^{2}})\right] $ \\ 
\  &  \\ 
$\theta _{3}(k)=$ & $-\ {\frac{1}{30\,{(1+4\,y)}^{7/2}}\ }\left[ \,48\,{k^{4}%
}+8\,{k^{2}}\,(17+80\,y+24\,{y^{2}})+3\,(31+224\,y+400\,{y^{2}})\right] $ \\ 
\  &  \\ 
$\theta _{4}(k)$ & $={\frac{16\,\sqrt{2}}{15\,({{2+5\,y)}^{5/2}}}}\,\ \left[
1536\,{k^{6}}\,{y^{2}}\,{{(1+2\,y)}^{2}}\,(1+4\,y+3\,{y^{2}})\right. $ \\ 
\  &  \\ 
\  & $-\ 3\,{{(2+5\,y)}^{3}}\,(434+2571\,y+4800\,{y^{2}}+2870\,{y^{3}})$ \\ 
\  &  \\ 
\  & $-\ 8\,{k^{2}}\,{{(2+5\,y)}^{2}}\,476+3774\,y+11043\,{y^{2}}+14075\,{%
y^{3}}+6780\,{y^{4}})$ \\ 
\  &  \\ 
\  & $-\ \left. 192\,{k^{4}}\,(28+344\,y+1697\,{y^{2}}+4268\,{y^{3}}+5691\,{%
y^{4}}+3727\,{y^{5}}+905\,{y^{6}})\right] $ \\ 
\  &  \\ 
$\theta _{5}(k)$ & $={\frac{1}{60\,(1+4\,y)^{7}}}\,\ \left[ 48\,{k^{4}}+8\,{%
k^{2}}\,(17+160\,y+368\,{y^{2}})+3\,{{(1+4\,y)}^{2}}\,(31+200\,y+496\,{y^{2}}%
)\right] $ \\ 
\  &  \\ 
$\theta _{6}(k)$ & $={\frac{1}{15\,(1+y)^{11/2}\,(1+4\,y)^{7/2}}\ }\left[
48\,{k^{4}}\,(7+46\,y+72\,{y^{2}})\right. $ \\ 
\  &  \\ 
\  & $+\ 8\,{k^{2}}\,(119+935\,y+2862\,{y^{2}}+3890\,{y^{3}}+1844\,{y^{4}})$
\\ 
\  &  \\ 
\  & $+\ \left. 3\,(1+y)^{2}\,(217+1486\,y+4236\,{y^{2}}+5944\,{y^{3}}+3472\,%
{y^{4}})\right] $%
\end{tabular}
\]

\subsubsection{\protect\bigskip Nucleus $^{40}Ca$}

Again, the expressions for $^{40}Ca$ have similar structure but are too long
to be presented here. They can be obtained in the form of user friendly
files for Mathematica 3.0 upon request.

\begin{quotation}
{\bf Figure Captions}

\bigskip

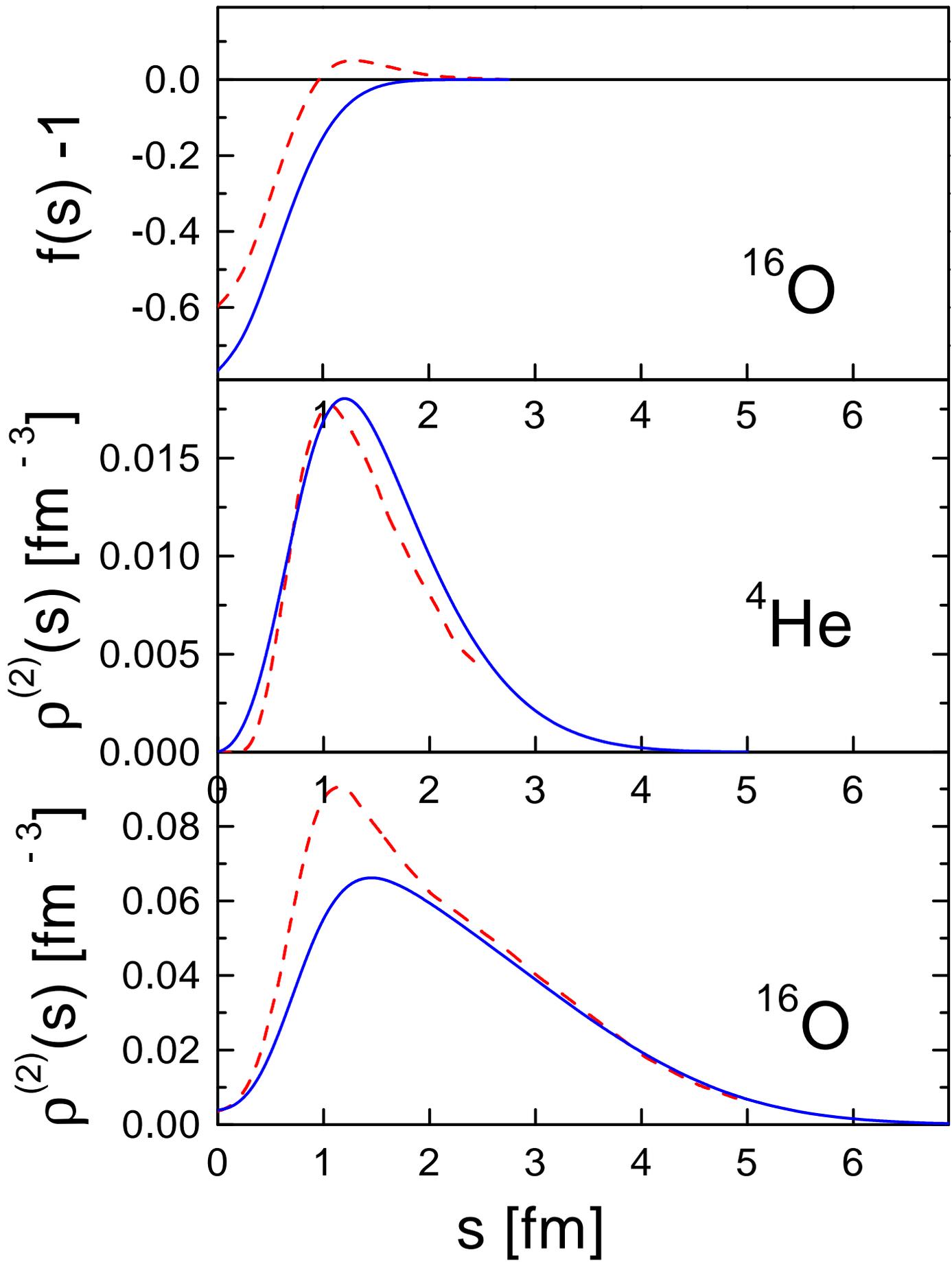
{\bf Fig. 1.} Comparison of the present results (solid lines) with the
Variational Monte-Carlo results for Argonne $v_{14}$ potential \cite{wsa} as
obtained in \cite{pwp},\cite{car} (dashed lines). Top panel - comparison of
the central correlation function. Next two panels - comparison of the
relative pair density distributions normalized as $4\pi \int \rho ^{\left(
2\right) }(s)s^{2}ds=1$ for $^{4}He$ and $^{16}O$, respectively.

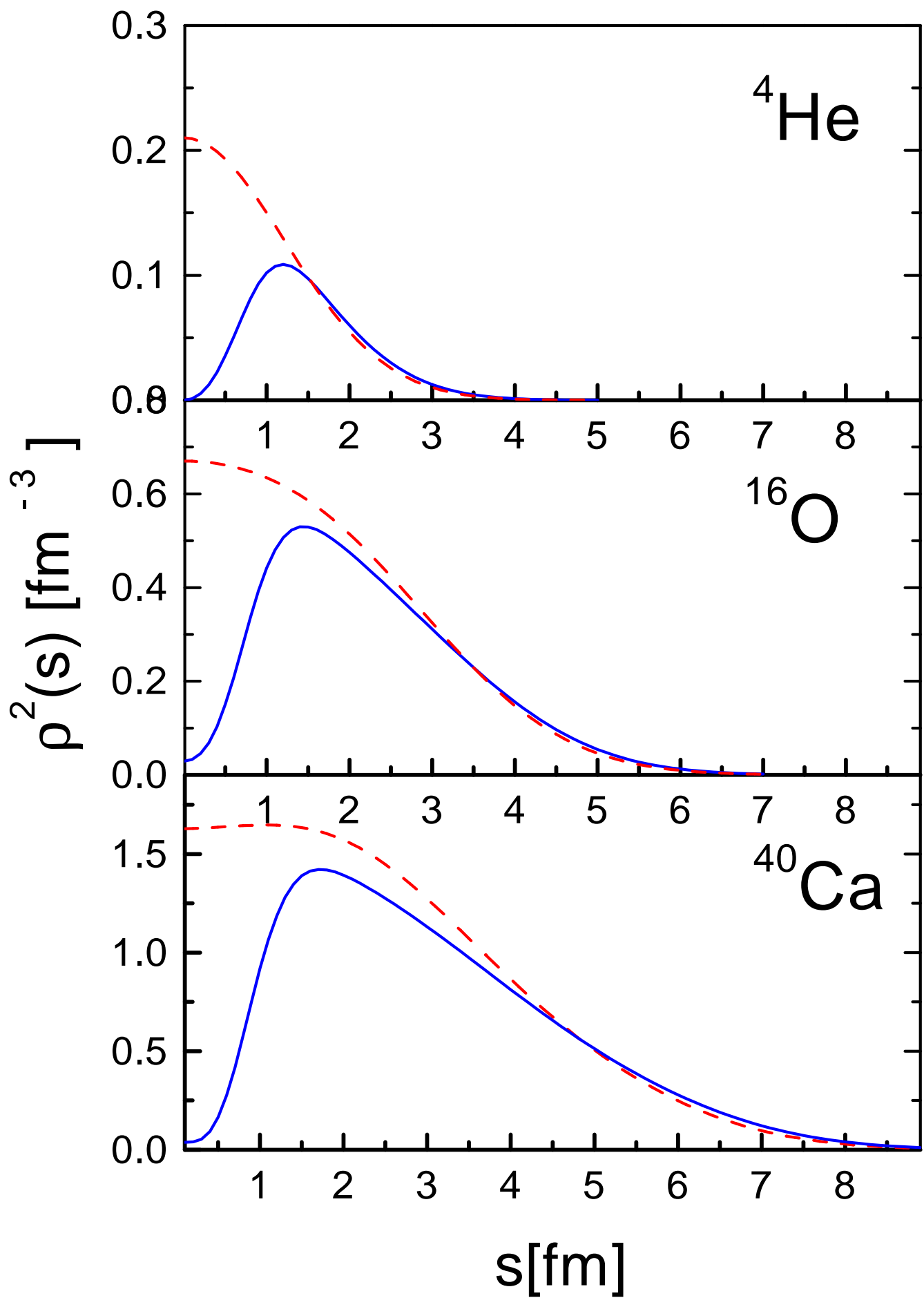
{\bf Fig. 2. }Comparison between correlated ($c\neq 0$ , solid lines) and
uncorrelated ($c=0$, dashed lines) results for the relative pair density
distribution (\ref{r2s}) normalized as $4\pi \int \rho ^{\left( 2\right)
}(s)s^{2}ds=1$ for $^{4}He$, $^{16}O$ and $^{40}Ca$.

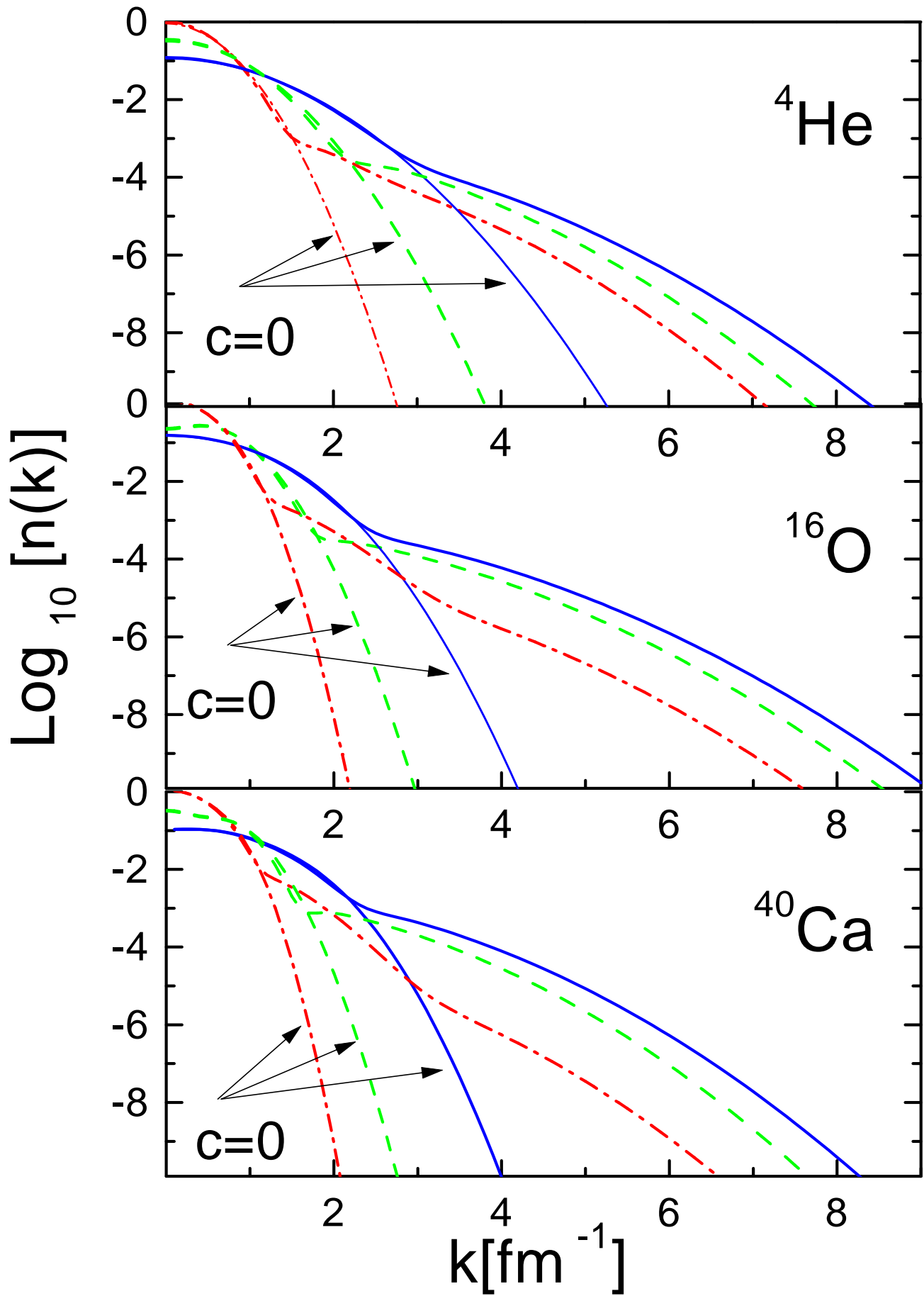
{\bf Fig. 3. }Correlated ($c\neq 0$) and uncorrelated ($c=0$) results for
the center-of-mass $n^{(2)}(K)$ (solid curve) and relative $n^{(2)}(k)$
(dashed-dot curve) pair momentum distributions and for the nucleon momentum
distribution $n(k)$ (dashed curve) for $^{4}He$ (top panel), $^{16}O$
(middle panel) and $^{40}Ca$ (bottom panel). All distributions are
normalized to unity as e.g. $4\pi \int n(k)k^{2}dk=1$.

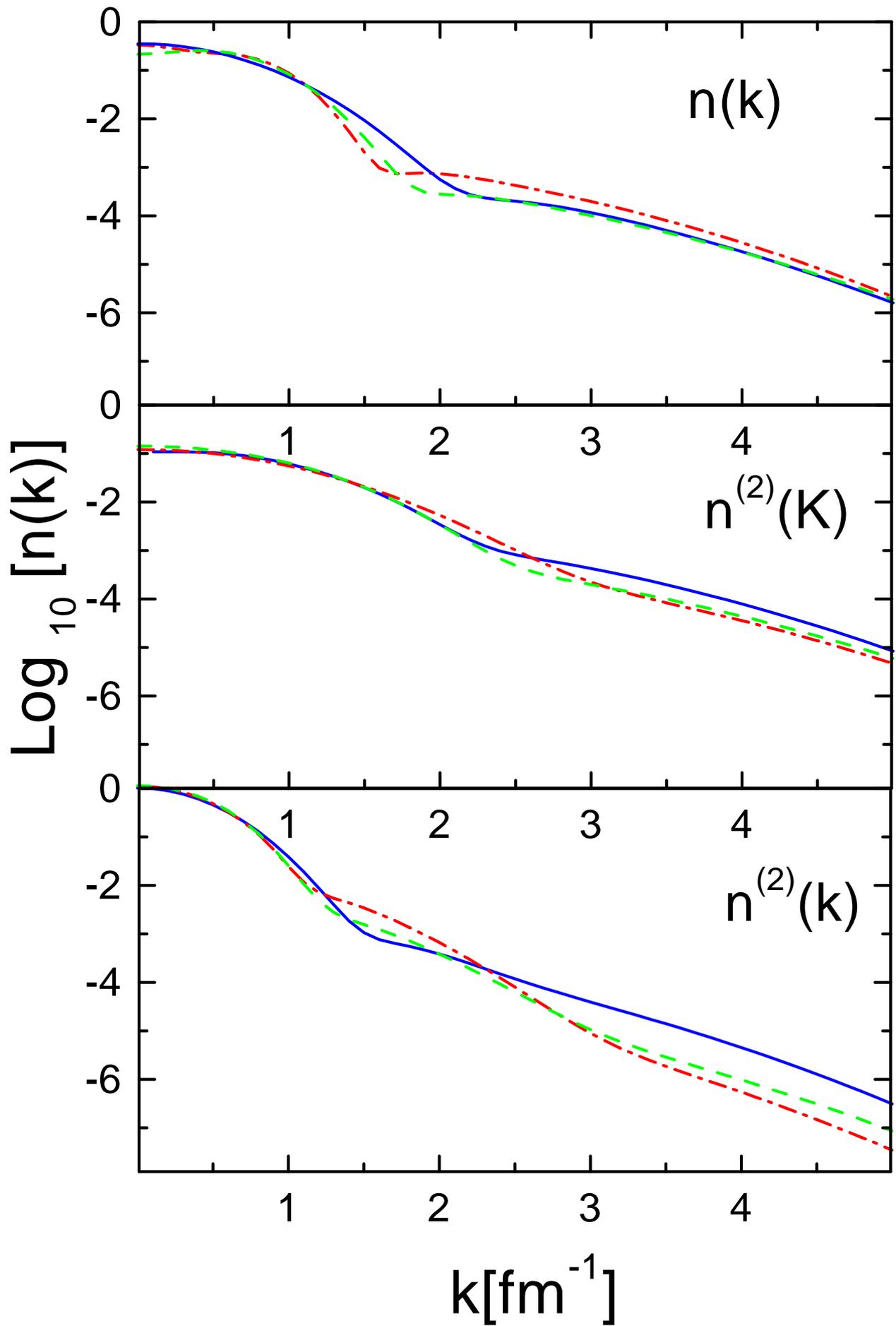
{\bf Fig. 4. }Correlated momentum distributions from Fig. 3, $n(k)$ (top
panel), $n^{(2)}(K)$ (middle panel) and $n^{(2)}(k)$ (bottom panel), but
collected for all three nuclei $^{4}He$ (solid curve), $^{16}O$ (dashed
curve) and $^{40}Ca$ (dashed-dot curve).
\end{quotation}

\end{document}